\documentclass[
    ,final            % use final for the camera ready runs
  ]
  {aipproc}

\layoutstyle{6x9}

\begin{document}

\title{Local temperature for dynamical black holes} 



\author{Sean A. Hayward}
{address={Center for Astrophysics, Shanghai Normal University, 100 Guilin Road, 
Shanghai 200234, China}} 

\author{R.~Di Criscienzo}
{address={Mc Lennan Physical Laboratories - Department of Physics,
  University of Toronto, 60 St. George Street, Toronto, ON, M5S 1A7,
  Canada}}

\author{M.~Nadalini}
  {address={Dipartimento di Fisica, Universit\`a di Trento and INFN, Gruppo 
  Collegato di Trento, Italia}} 

\author{L.~Vanzo}
  {address={Dipartimento di Fisica, Universit\`a di Trento and INFN, Gruppo 
  Collegato di Trento, Italia}} 

\author{S.~Zerbini}
  {address={Dipartimento di Fisica, Universit\`a di Trento and INFN, Gruppo 
  Collegato di Trento, Italia}} 

\begin{abstract} 
A local Hawking temperature was recently derived for any future outer trapping 
horizon in spherical symmetry, using a Hamilton-Jacobi tunneling method, and is 
given by a dynamical surface gravity as defined geometrically. Descriptions are 
given of the operational meaning of the temperature, in terms of what observers 
measure, and its relation to the usual Hawking temperature for static black 
holes. Implications for the final fate of an evaporating black hole are 
discussed. 
\end{abstract}

\maketitle 

%%%%%%%%%%%%%%%%%%%%%%%%%%%%%%%%%%%%%%%%%%%%
%% MAINMATTER
%%%%%%%%%%%%%%%%%%%%%%%%%%%%%%%%%%%%%%%%%%%%

\section{0. Introduction}

Hawking \cite{Haw} showed that stationary black holes radiate thermally at a 
temperature given by their surface gravity. In a quasi-stationary (or 
adiabatic) approximation, a radiating black hole loses energy and therefore 
shrinks. The rate accelerates. This raises the question of the final fate of 
evaporation, including the supposed information paradox. 

The fundamental problem is that an evaporating black hole is non-stationary, 
while the classic derivations of Hawking temperature do not obviously 
generalize beyond stationary black holes. So the question arises: is there in 
any sense a Hawking temperature for dynamical black holes?  

Traditionally, black holes have generally been defined by event horizons 
\cite{HE,Wal}, despite their physically unlocatable nature, leading to some 
confusion that they may be the source of Hawking radiation. Fortunately recent 
years have seen the development of a local theory of dynamical black holes, 
based on a refinement of apparent horizons, trapping horizons \cite{bhd,1st}, 
which have physical properties such as mass and surface gravity, satisfying 
physically interpretable equations \cite{AK,bhd2}. This theory is practical 
enough to apply to violent astrophysical processes such as binary black-hole 
mergers \cite{bhd6}, which may be observable in the near future via 
gravitational-wave detectors. 

Contemporaneously, Parikh \& Wilczek \cite{PW} developed a tunneling method to 
derive temperature for stationary black holes, making precise the intuitive 
idea of Hawking radiation in terms of pair production. A Hamilton-Jacobi 
variant turns out to work even for dynamical black holes \cite{bht}, yielding a 
local temperature precisely for future outer trapping horizons, which were 
previously proposed as a local definition of black holes as part of the above 
theory \cite{bhd}. Moreover, the temperature is given by the surface gravity as 
previously defined for dynamical black holes on geometrical grounds \cite{1st}. 

The article is organized as follows. 

\begin{enumerate} 

\item Geometry of dynamical black holes: trapping horizons, area, mass,     
    surface gravity
    
\item Hamilton-Jacobi tunneling method: local temperature 

\item Operational meaning: redshift and observed temperature 

\item Static, asymptotically flat space-times: surface gravity vs.\ Killing 
    ``surface gravity'', local temperature vs. Hawking temperature 
    
\item Extremal limits: charged stringy black hole 

\item Remarks: evaporation and final fate 
    
\end{enumerate}
    
General Relativity is assumed throughout, but not the Einstein equation with 
prescribed source, so any semi-classical model is included. 

\section{1. Geometry of dynamical black holes}
    
Spherical symmetry will be assumed throughout, with spheres of area $A$. The 
area radius $r=\sqrt{A/4\pi}$ is convenient. A sphere is {\em untrapped}, {\em 
marginal} or {\em trapped} if $g^{-1}(dr)$ is respectively spatial, null or 
temporal, and {\em future} or {\em past} trapped or marginal if $g^{-1}(dr)$ is 
respectively future or past causal \cite{1st}. A hypersurface foliated by 
marginal spheres is a {\em trapping horizon} \cite{bhd}. 

The active gravitational mass $m$ \cite{MS} is defined by 
\begin{equation} 
1-2m/r=g^{-1}(dr,dr) 
\end{equation} 
in units $G=1$, where spatial metrics are positive definite. It has various 
physical or mathematically useful properties \cite{sph,inq}, of which the key 
one here is that a sphere is trapped, marginal or untrapped if respectively 
$r<2m$, $r=2m$ or $r>2m$.  

There is a preferred time vector $K=g^{-1}({*}dr)$ identified by Kodama 
\cite{Kod}, where $*$ is the Hodge operator in the space normal to the spheres 
of symmetry: 
\begin{equation} 
K\cdot dr=0,\qquad 
g(K,K)=-g^{-1}(dr,dr). 
\end{equation} 
Then both $K$ and the energy-momentum density with respect to it are 
divergence-free, and the Noether charge of the latter is $m$. The Kodama vector 
coincides with the static Killing vector of standard black holes such as 
Schwarzschild and Reissner-Nordstr\"om. Note that $K$ is temporal, null or 
spatial respectively on untrapped, marginal or trapped spheres. 

Surface gravity was defined by \cite{1st}
\begin{equation}
\kappa={*}d{*}dr/2 
\end{equation}
where $d$ is the exterior derivative in the normal space, i.e. ${*}d{*}d$ is 
the wave operator in the normal space. It also has various physical or 
mathematically useful properties, of which the key one here is that  
\begin{equation} 
K^a\nabla_{[b}K_{a]}\cong\pm\kappa K_b 
\end{equation} 
where $\cong$ denotes evaluation on a trapping horizon $r\cong2m$, similarly to 
the usual Killing identity. Then a trapping horizon is {\em outer}, {\em 
degenerate} or {\em inner} respectively if $\kappa>0$, $\kappa=0$ or $\kappa<0$ 
\cite{1st}. Examples of all types are provided by Reissner-Nordstr\"om 
solutions: the future or past trapping horizons are respectively the Killing 
horizons of the black or white hole, being outer, degenerate or inner as 
appropriate. In vacuo, $\kappa=m/r^2$ \cite{1st}, therefore reducing to the 
Newtonian surface gravity in the Newtonian limit, since $m$ reduces to the 
Newtonian mass. Thus $\kappa$ also provides a relativistic definition of the 
surface gravity of planets and stars. 

For an advanced time $v$, the generalized advanced Eddington-Finkelstein metric 
\begin{equation} 
ds^2=r^2d\Omega^2+2e^{\Psi}dvdr-e^{2\Psi}Cdv^2
\end{equation}
with $(C,\Psi)$ functions of $(r,v)$ and
\begin{equation} 
C=1-2m/r
\end{equation}
is valid \cite{bht} in untrapped regions, $C>0$, on future marginal surfaces, 
$C=0$, and in future trapped regions, $C<0$, as appropriate for black holes 
rather than white holes. Note that $C$ is an invariant, but $\Psi$ is not, due 
to the freedom $v\to\tilde{v}(v)$. Also $K=e^{-\Psi}\partial_v$ and 
$\kappa=e^{-\Psi}\partial_r(e^{\Psi}C)/2$, so 
\begin{equation} 
\kappa\cong\partial_r C/2.
\end{equation}

\section{2. Hamilton-Jacobi tunneling method}

The WKB approximation of the tunneling probability $\Gamma$ along the 
classically forbidden trajectory from inside to outside the horizon is 
\begin{equation} 
\Gamma\propto\exp(-2\Im I)
\end{equation}
in units $\hbar=1$, where $\Im I$ is the imaginary part of the action $I$ on 
the classical trajectory. For a massless scalar field $\phi=\phi_0\exp(iI)$ in 
the eikonal (or geometrical optics) approximation, the amplitude $\phi_0$ is 
slowly varying and the action 
\begin{equation} 
I=\int\omega e^\Psi dv-\int kdr
\end{equation}
is rapidly varying, defining angular frequency $\omega$ and wave number $k$,  
where $e^{\Psi}$  is included to make $\omega$ and $I$ invariant, recalling the 
freedom $v\to\tilde{v}(v)$. Equivalently, $\omega=K\cdot 
dI=e^{-\Psi}\partial_vI$, $k=-\partial_r I$. Then the wave equation 
$\nabla^2\phi=0$ yields the Hamilton-Jacobi equation 
\begin{equation} 
g^{-1}(\nabla I,\nabla I)=0
\end{equation}
which becomes
\begin{equation} 
2\omega k-Ck^2=0.
\end{equation}
Then $k=0$ yields the ingoing modes, while $k=2\omega/C$ yields the outgoing 
modes. Since $C\cong 0$ at a trapping horizon $r\cong r_0$, $I$ has a pole, 
evaluated by $C\approx(r-r_0)\partial_r C$. Thus $k\approx\omega/\kappa(r-r_0)$ 
if $\kappa\not\cong0$, yielding
\begin{equation} 
\Im I\cong\pi\omega/\kappa.
\end{equation}
Then the tunneling probability takes a thermal form 
\begin{equation}
\Gamma\propto\exp(-\omega/T) 
\end{equation}
with temperature $T$ given by 
\begin{equation}
T\cong\kappa/2\pi.
\end{equation}
For this to be positive, $\kappa>0$, so the trapping horizon is of the outer 
type. Thus the method has derived a positive temperature if and only if there 
is a future outer trapping horizon, remarkably confirming the local definition 
of black hole \cite{bhd,1st}. 

Note that this is nothing to do with event horizons. There may be no event 
horizon in the space-time. If there is, and it does not coincide with a 
trapping horizon, the above method does not yield a thermal spectrum on it. 
Generally, the method gives no reason to expect a thermal spectrum everywhere 
in the space-time, including at infinity, but only on a future outer trapping 
horizon, and therefore approximately in a neighbourhood. 

\section{3. Operational meaning}

The integral curves of $K$, outside the horizon, are the worldlines of 
preferred observers, who would be static observers in the static case. Their 
velocity vector is $\hat K=K/\sqrt{C}$. The angular frequency measured by such 
observers is $\hat\omega=\hat K\cdot dI=\omega/\sqrt{C}$. Such observers 
therefore measure a thermal spectrum with temperature 
\begin{equation} 
\hat T\approx T/\sqrt{C}
\end{equation}
to leading order near the horizon. The invariant redshift factor $\sqrt{C}$ is 
familiar from the Schwarzschild case, where it reflects the acceleration 
required to keep an observer static \cite{BD}. So this is the operational 
meaning of $T$: not that someone is measuring $T$ directly, but that the 
preferred observers just outside the horizon measure $T/\sqrt{C}$, which 
diverges at the horizon. Then $T$ itself can be interpreted as a 
redshift-renormalized temperature, finite at the horizon. One might also 
conjecture that freely falling observers crossing the horizon measure a 
temperature of the order of $T$, as predicted for static cases \cite{BD}. 

\section{4. Static, asymptotically flat space-times}

The surface gravity $\kappa$ coincides with the usual definition of the Killing 
``surface gravity'' $\kappa_\infty$ for standard static black holes such as 
Schwarzschild and Reissner-Nordstr\"om. However, it does not coincide if 
$\Psi\not\cong0$, requiring further explanation. 

Static metrics can be written as
\begin{equation}
ds^2=r^2d\Omega^2+C^{-1}dr^2-Ce^{2\Psi}dt^2 
\end{equation}
where $(C,\Psi)$ are henceforth functions of $r$ alone, the notation being 
consistent with the above. The static Killing vector $\partial_t$ is 
\begin{equation}
K_\infty=e^\Psi K.
\end{equation}
Then $\kappa_\infty$ is defined by $K_\infty^a\nabla_bK_{\infty 
a}\cong\kappa_{\infty}K_{\infty b}$, yielding
\begin{equation} 
\kappa_{\infty}\cong e^{\Psi}\kappa.
\end{equation}
This discrepancy can be understood as follows. A textbook method derives the 
gravitational redshift of light along a given ray \cite{Wal}: 
$\sqrt{-g(\partial_t,\partial_t)}\hat\omega=e^\Psi\sqrt{C}\hat\omega$ is 
constant along the ray. If the space-time is asymptotically flat, with $(t,r)$ 
being Minkowski coordinates as $r\to\infty$, then $C\to1$, $\Psi\to0$ and 
$\partial_t\to K$. Note that it is precisely here where the generally 
non-invariant $\Psi$ acquires a specific meaning. Then the angular frequency 
measured by static observers at infinity is 
\begin{equation}
\omega_\infty=e^\Psi\sqrt{C}\,\hat\omega 
\end{equation}
and the corresponding temperature measured by such observers is
\begin{equation}
T_\infty=e^\Psi\sqrt{C}\,\hat T 
\end{equation}
which is the Tolman relation \cite{Tol}. Thus
\begin{equation} 
T_\infty\cong e^\Psi T
\end{equation}
which indeed corresponds to
\begin{equation}
\kappa_\infty\cong2\pi 
T_\infty 
\end{equation}
Hence $e^{\Psi}$ appears as a relative redshift between the horizon and 
infinity. The Tolman relation mixes the redshift factors, $\sqrt{C}$ invariant 
and $e^\Psi$ relative. 

So the appropriate local temperature at the horizon is $T$ and generally not 
$T_\infty$ even in the static case. Likewise, the local surface gravity is 
$\kappa$ and generally not the textbook definition $\kappa_{\infty}$. Recall 
that the physical interpretation of $\kappa_\infty$ is the force at infinity 
per unit mass required to suspend an object from a massless rope just outside 
the horizon \cite{Wal}. This ``surface gravity at infinity'' would seem to be 
an oxymoron. It bears no relation to how Newtonian surface gravity is defined, 
whereas $\kappa$ reduces as above to the Newtonian surface gravity in vacuo. 

The relative redshift factor stems from using the Kodama vector $K$ instead of 
$\partial_t$, since the latter does not exist in dynamic cases. Thus one can 
deal in a unified way with such situations as an accreting black hole which 
settles down to a static state, or a static black hole which starts to 
evaporate. 

\section{5. Extremal limits}

Lest there still be doubts about the above unorthodox conclusion, a key 
property of surface gravity is that it should vanish in extremal cases. A good 
example is provided by charged stringy black holes, which are non-vacuum 
solutions of Einstein-Maxwell dilaton gravity in the string frame 
\cite{Gib,GHS}: 
\begin{equation} 
ds^2=r^2d \Omega^2+\frac{dr^2}{\left(1-a/r\right)\left(1-b/r\right)}-\left( 
\frac{1-a/r}{1-b/r}\right)dt^2
\end{equation}
where $a>b>0$. The horizon radius is $r\cong a$. 

The extremal limit as defined by global structure is $b\rightarrow a$. The 
Killing ``surface gravity'' $\kappa_{\infty}\cong1/2a$ does not vanish in this 
limit, whereas extremal black holes are expected to be zero-temperature 
objects. Remarkably, $\kappa\cong(a-b)/2a^2$ vanishes in the extremal limit. 
This is striking confirmation of the appropriateness of $\kappa$ over 
$\kappa_\infty$ as a local surface gravity. 

\section{6. Remarks}

Returning to the main result: dynamical black holes indeed possess a local 
temperature $T$, with the operational meaning that it determines the redshifted 
temperature $T/\sqrt{1-2m/r}$ measured by Kodama observers just outside a 
trapping horizon. The method works precisely for future outer trapping 
horizons, proposed previously to define black holes on purely geometrical 
grounds, and $T=\kappa/2\pi$ in terms of the geometrically defined surface 
gravity $\kappa$. This confirms the quasi-stationary picture of black-hole 
evaporation in early stages. 

Apart from the restriction to spherical symmetry, the derivation is general, 
exact, simple, independent of model or semi-classical ambiguities, and 
therefore robust. It yields a clear conclusion on the much debated issue of 
whether Hawking radiation is a mysterious global effect associated with event 
horizons, or even the entire space-time, or a local geometrical effect. 

The result holds formally for arbitrarily fast evaporation, even in regimes 
where one normally expects a semi-classical approximation to break down. With 
this qualification, it strongly suggests that evaporation proceeds until 
$\kappa\to0$. While this is reminiscent of quasi-stationary arguments, it has a 
different meaning, since $\kappa$ is generally not the surface gravity of a 
static black hole with the same mass and whatever other parameters in a given 
model. 

A common idea is that evaporation results in an extremal remnant \cite{Gid,DL}. 
For instance, an outer ($\kappa>0$) and inner ($\kappa<0$) trapping horizon 
might asymptote to the same null hypersurface, effectively forming a degenerate 
($\kappa=0$) trapping horizon. Another idea is that the outer and inner 
trapping horizons merge smoothly at a single moment of extremality where 
$\kappa$ vanishes \cite{02bh20}. The results here are consistent with either 
picture. 

\begin{theacknowledgments}
SAH thanks Ted Jacobsen and Alex Nielsen for discussions. SAH was supported by 
the National Natural Science Foundation of China under grants 10375081, 
10473007 and 10771140, by Shanghai Municipal Education Commission under grant 
06DZ111, and by Shanghai Normal University under grant PL609. RDC wishes to 
thank the INFN - Gruppo Collegato di Trento and the Department of Physics at 
the University of Trento, where part of this work has been done. 
\end{theacknowledgments}

\bibliographystyle{aipproc}

\end{document}